\documentclass[aps,prd,showpacs,showkeys,floatfix,preprintnumbers,
footinbib]{revtex4}
\usepackage[utf8]{inputenc}
\usepackage{graphicx}
\usepackage{dcolumn}
\usepackage{bm}
\usepackage{amssymb}
\usepackage{amsmath}
\usepackage{subfigure}
\usepackage{tabularx}
\usepackage{hyperref}
\usepackage{color}
\usepackage{multirow,array}
\newcolumntype{P}[1]{>{\centering\arraybackslash}p{#1}}

\begin{document}
\title{
Finite temperature properties of a modified
Polyakov$-$Nambu$-$Jona-Lasinio model}

\author{Abhijit Bhattacharyya}
\email{abphy@caluniv.ac.in}
\affiliation{Department of Physics, University of Calcutta,
             92, A.P.C Road, Kolkata-700009, INDIA}
\author{Paramita Deb}
\email{paramita.deb83@gmail.com}
\affiliation{Indian Institute of Technology Bombay, Mumbai-400076, INDIA}
\author{Sanjay K. Ghosh}
\email{sanjay@jcbose.ac.in}
\author{Soumitra Maity}
\email{soumitra.maity1984@gmail.com}
\author{Sibaji Raha}
\email{sibaji@jcbose.ac.in}
\author{Rajarshi Ray}
\email{rajarshi@jcbose.ac.in}
\affiliation{Center for Astroparticle Physics \&
Space Science, Block-EN, Sector-V, Salt Lake, Kolkata-700091, INDIA 
 \\ \& \\ 
Department of Physics, Bose Institute, \\
93/1, A. P. C Road, Kolkata - 700009, INDIA}
\author{Kinkar Saha}
\email{saha.k.09@gmail.com}
\affiliation{Department of Physics, University of Calcutta,
             92, A.P.C Road, Kolkata-700009, INDIA}
\author{Sudipa Upadhaya}
\email{sudipa.09@gmail.com}
\affiliation{Variable Energy Cyclotron Centre,
             1/AF, Bidhannagar, Kolkata-700064, INDIA}

\begin{abstract}
Thermodynamic properties of strongly interacting matter are investigated
using the Polyakov loop enhanced Nambu$-$Jona-Lasinio model along with
some modifications to include the hadrons. Various observables are shown
to have a close agreement with the numerical data of QCD on lattice.
The advantage of the present scheme over a similar study using a
switching function is that here no extra parameters are to be fitted.
As a result the present scheme can be easily extended for finite
chemical potentials.

\end{abstract}
\maketitle

\section{INTRODUCTION}

The programme of various ongoing and upcoming relativistic heavy-ion
collision experiments is to explore the properties of strongly
interacting matter at finite temperatures and densities. There has been
major theoretical advances in the finite temperature properties using
simulations of Quantum Chromodynamics (QCD) on space-time lattices.
Currently the temperature $T_c$ where a rapid cross-over from hadronic
to partonic matter takes place is found to be close to the pion mass.
The reported values are 155 MeV~\cite{Bazavov12, Bazavov14} and 150 MeV
\cite{Borsanyi14} from HotQCD and Wuppertal-Budapest (WuB)
collaborations respectively. At the same time various QCD inspired model
frameworks have been developed to extract interesting physical insights.
The Nambu-Jona-Lasinio or the NJL model \cite{YNambu, Hatsuda2, Vogl,
Klevansky, Hatsuda1, Buballa1} is one such model which effectively
explains key features of QCD like chiral symmetry breaking and its
restoration. However the effects of the gluon degrees of freedom are not
adequately addressed in such models.  The extensions like the Polyakov
loop enhanced Nambu-Jona-Lasinio (PNJL) model \cite{Fukushima, pisarski,
Ratti, fuku2, hansen} encapsulates this missing feature by including a
temporal background gluon field. As a result, both chiral and
deconfinement aspects are captured within a single framework. Various
studies of the basic thermodynamic variables computed in the mean field
framework display strong similarity to lattice results
(see~\cite{hybrid} and references therein). 

With more sophisticated techniques and higher computational power, some
of the observables from lattice QCD have been extrapolated to the
physical continuum limit~\cite{Bazavov14, Borsanyi14}. Many of these
results are quite different from those obtained earlier on smaller
lattices. In view of this a reparametrization of the PNJL model was
done~\cite{repara} to obtain a quantitative agreement with the lattice
QCD data. One important lacunae observed in this study is the mismatch
of the results for temperatures close to or below $T_c$. The reason was
identified as the absence of hadronic contribution in the PNJL model.
Several other attempts were going on to construct a suitable model to
match the lattice data. In HRG+chiral perturbation theory \cite{fodor1},
below the transition temperature, the decrease of the absolute value of
chiral condensate is well described. Also HRG model is able to
reproduce LQCD data on temperature dependence of the Polyakov-loop
itself \cite{megias}. Similarly, a quark-hadron hybrid
model~\cite{miyahara} has been constructed by taking quark and hadron
contributions simultaneously. The hadron volume fraction function is
used to switch from one phase to other and hadron-quark transition
temperature is defined in the view of the ratio of quark and hadron
contribution. In this direction some of us studied a hybrid model by
coupling the HRG model and the PNJL model via a switching
function~\cite{hybrid}. 

Here we study an alternative scheme where the hadron contributions are
added in a simple way except that we consider their medium dependent
masses. The confinement feature through the Polyakov loop always
suppresses the contribution of the constituent quarks at low
temperatures and densities. The switching function was necessary to
rather cut off the contribution of the hadrons at high temperatures
and densities. Instead of the switching function, here the rising
effective masses of the hadrons will naturally make them unfavorable in
the thermodynamics.

In the next section, we briefly outline the PNJL model. The following
section gives a description of how we handle the hadrons. This is
followed by our results and conclusions.

\section{PNJL Model}
\label{sc.pnjl}

We now discuss the particular form of the PNJL model as discussed in
Ref.~\cite{repara}, which will be employed here. The scheme in the PNJL
model was to add a Polyakov loop effective potential to the NJL
model~\cite{Ogilvie, Fukushima, Ratti}. The chiral properties are taken
care of by the NJL part, while the confinement properties and the
gluonic contributions are effectively incorporated through the Polyakov
loop potential.  Various studies have been carried out using PNJL model
with 2 and 2+1 flavors~\cite{Ratti, Ray, Mukherjee, Ghosh, ciminale,
Bhattacharyya, Shao, Tang, Haque, Peixoto, PNJL-reg}.  For our study we
shall use the 2+1 flavor model having up to six quark interactions.  The
thermodynamic potential is given as~\cite{repara},

\small
\begin{eqnarray}
\Omega (\Phi,\bar{\Phi},\sigma_f,T,\mu ) &=&
2g_S\sum_{f=u,d,s}\sigma_f^2 -
\frac{g_D}{2}\sigma_u\sigma_d\sigma_s
- 6\sum_f\int_0^\infty\frac{d^3p}{(2\pi)^3}E_f
\Theta(\Lambda-|\vec{p}|)
\nonumber \\
&& - ~2T\sum_f\int_0^\infty\frac{d^3p}{(2\pi)^3} \ln{\left[ 
1+3\left( \Phi+\bar{\Phi} e^{-\left( E_f-\mu_f \right) /T} \right)
e^{-\left( E_f-\mu_f \right) /T } + 
e^{-3\left( E_f-\mu_f \right) /T } \right]  }
\nonumber \\
&& - ~2T\sum_f\int_0^\infty\frac{d^3p}{(2\pi)^3} \ln{\left[ 
1+3\left( \bar{\Phi}+\Phi e^{-\left( E_f+\mu_f \right) /T} \right)
e^{-\left( E_f+\mu_f \right) /T } + 
e^{-3\left( E_f+\mu_f \right) /T } \right]  }
\nonumber \\
&&
+ ~\mathcal{U'}(\Phi,\bar{\Phi},T).
\label{eq.Potential}
\end{eqnarray}
\normalsize

\noindent
The first five terms on the R.H.S. are the terms of the NJL model
suitably modified due to the Polyakov loop.  Here
$\sigma_f=\langle\bar\psi_f\psi_f\rangle$ correspond to the two light
quark ($f=u,d$) condensates and the strange ($f=s$) quark condensate
respectively. There is a four quark coupling term with coefficient $g_S$
and a six quark coupling term breaking the axial U(1) symmetry
explicitly with a coefficient $g_D$.  The corresponding quasiparticle
energy is $E_f=\sqrt{p^2+M_f^2}$, for a given flavor $f$.  The
dynamically generated constituent quark masses is given by,

\begin{equation}
M_f=m_f-2g_S\sigma_f+\frac{g_D}{2}\sigma_{f+1}\sigma_{f+2}
,
\end{equation} 

\noindent
where, if $\sigma_f=\sigma_u$, then $\sigma_{f+1}=\sigma_d$ and
$\sigma_{f+2}=\sigma_s$, and so on in a clockwise manner. 

\noindent
The third term on the R.H.S. of Eq.~(\ref{eq.Potential}) gives the zero
point energy, while the fourth and fifth terms are the finite
temperature and chemical potential contributions of the constituent
quarks and anti-quarks respectively. The latter two terms arise from the
fermion determinant in the NJL model duly modified by the fields
corresponding to the traces of Polyakov loop and its conjugate given by
$\Phi=\frac{Tr_cL}{N_c}$ and $\bar{\Phi}=\frac{Tr_cL^{\dagger}}{N_c}$
respectively. Here $L(\vec{x})=\mathcal{P} exp\left[ i\int_0^{1/T}d\tau
A_4(\vec{x},\tau) \right]$ is the Polyakov loop, and $A_4$ is the
temporal component of background gluon field.

\par
The effective potential for the $\Phi$ and $\bar{\Phi}$ fields are given
by $\mathcal{U'}$, appearing as the last term in
Eq.~(\ref{eq.Potential}). Various forms of the potential exist in the
literature (see e.g.  ~\cite{Robner, Fuku, Ghosh, Contrera, Qin}). We
shall use the form prescribed in~\cite{repara} which reads as,

\begin{equation}
\frac{\mathcal{U'}(\Phi,\bar{\Phi},T)}{T^4}=
\frac{\mathcal{U}(\Phi,\bar{\Phi},T)}{T^4}-\kappa
ln[J(\Phi,\bar{\Phi})].
\label{eq.Ppotential}
\end{equation}

\noindent
Here ${\mathcal U}(\Phi,\bar{\Phi},T)$ chosen as a Landau-Ginzburg type
potential commensurate with the global Z(3) symmetry of the Polyakov
loop is given as~\cite{Ratti},

\begin{equation}
\frac{\mathcal{U}(\Phi,\bar{\Phi},T)}{T^4}=
-\frac{b_2(T)}{2}\bar{\Phi}\Phi-\frac{b_3}{6}
(\Phi^3+\bar{\Phi}^3)+\frac{b_4}{4}(\bar{\Phi}\Phi)^2
\end{equation}

\noindent
The coefficient $b_2(T)$ is chosen to have a temperature dependence of
the form~\cite{repara},

\begin{equation}
b_2(T)=a_0+a_1exp(-a_2\frac{T}{T_0})\frac{T_0}{T},
\end{equation}

\noindent
and $b_3$ and $b_4$ are chosen to be constants. The term
$J[\Phi,\bar{\Phi}] = (1 - 6\,\bar{\Phi} \Phi + 4\,(\bar{\Phi}^3 +
\Phi^3) - 3\,(\bar{\Phi} \Phi)^2)$ is the Jacobian of transformation
from the Polyakov loop to its traces. $\kappa$ is a dimensionless
parameter which is determined phenomenologically.

\begin{table}[!htb]
\begin{tabular}{|c|c|c|c|c|}
\hline
\hline
$m_u$ (MeV) & $m_s$ (MeV) & $\Lambda$ (MeV) &
$g_s\Lambda^2$ & $g_D\Lambda^5$ \\
\hline
\hline
5.5 & 134.758 & 631.357 & 3.664 & 74.636 \\
\hline
\hline
\end{tabular}
\caption{Parameters in the NJL model}
\label{tb.njlpara}
\end{table}

\begin{table}[!htb]
\begin{tabular}{|c|c|c|c|c|c|c|}
\hline
\hline
$T_0$ (MeV) & $a_0$ & $a_1$ & $a_2$ & $b_3$ & $b_4$ &
$\kappa$\\
\hline
\hline
175 & 6.75 & -9.0 & 0.25 & 0.805 & 7.555 & 0.1 \\
\hline \hline
\end{tabular}
\caption{Parameters for the Polyakov loop potential.}
\label{tb.polpara}
\end{table}

The different parameter values in the NJL terms are given in
Table~\ref{tb.njlpara}.  And the parameter values used in the Polyakov
loop potential are given in Table~\ref{tb.polpara}.

\par
Previously some of us~\cite{repara} discussed that this model gives a
crossover temperature of $T_c\sim160~{\rm MeV}$ as well as quantitative
agreement of temperature variations of pressure and various other
observables commensurate with the observations in lattice QCD in the
continuum limit~\cite{repara}. However the quantitative agreement though
close, was not exact in different ranges of temperatures. Significant
discrepancies appeared in the low temperature region where the hadronic
degrees of freedom dominate. A possible step towards removal of this
lacunae was proposed by us~\cite{hybrid} by coupling the PNJL model
with the Hadron Resonance Gas model via a switching function. This
scheme was successful in getting a much better agreement between the
results from PNJL model and the lattice QCD data. Here a key role is
played by the switching function that switches the hadronic or the
partonic degrees of freedom. However this approach requires us to
immaculately choose a form and parametrization of the switching function
itself. Here we discuss an alternative scheme where PNJL model is modified
such that the hadronic contributions would appear more naturally in the
relevant region of the phase space and shut off in other regions,
without having to use a switching by hand. In the next section we shall
describe this scheme.

\section{Hadronic Sector}

Our aim is to include all the correct degrees of freedom allowed in
strong interactions in our model framework. As discussed
in~\cite{hybrid} (and references therein), the prominent degrees of
freedom would depend on the thermodynamic conditions. This gave a scope
for introducing the phenomenologically determined switching function
in~\cite{hybrid}, and couple the PNJL model to the HRG model. Here we
ask if a more natural mechanism exists to include the hadronic
contributions. 

As is well known, the thermodynamic potential given by
Eq.~\ref{eq.Potential} is obtained in the mean field approximation for
the quark propagators. A consistent method to extract the thermodynamic
potential beyond mean field for a quark meson plasma in the framework of
the NJL model was outlined in~\cite{bymn1-klev, bymn2-klev, bymn3-blas}.
The mesonic contributions appear in the next to leading order
contributions in a $1/N_c$ expansion in the form of ring diagrams.  For
a meson $M$ the contribution to the thermodynamic potential is given by,

\begin{equation}
 \delta\Omega_M=g_M \int \frac{d^3p}{(2\pi)^3} \int d\omega
\left[\frac{\omega}{2} + T~{\rm ln}(1-e^{-\frac{\omega}{T}})\right]
\frac{1}{\pi}\frac{d\delta_M(\omega,\vec{p},T)}{d\omega}.
\end{equation}
Here, $g_M$ is the internal degrees of freedom of the meson and
$\delta_M(\omega,\vec{p},T)$ is the scattering phase shift of a quark
and anti-quark in the $M$ channel.

Extensions of this work in the PNJL model has been done
in~\cite{bym1-2008, bym2-2011, bym3-2014, bym4-blas, bym5-2017}, wherein
the authors have studied various effects of this additional contribution
to the mean-field thermodynamic potential. On the other hand here we set
out to make a detailed study of the various thermodynamic observables
and contrast them to the results reported in the continuum limit in the
lattice QCD framework. Here we do not try to be rigorous with the beyond
mean-field calculations but carry out a simple heuristic approach.  We
simply add the hadronic contribution to the mean field PNJL model. The
masses of such hadronic excitations may be computed from the pole
condition in the respective polarizations, and would therefore depend
implicitly on the mean fields and explicitly on the thermodynamic
parameters. This approach is similar to the near-pole approximation
$(\omega^2=E_M^2=\vec{p}^2+m_M^2)$ of the above thermodynamic
potential~\cite{bymn1-klev}.  In practice this approach is similar to our
earlier approach~\cite{hybrid} of adding the hadronic contribution to
the PNJL model, but without a switching function. Here the effect of
switching off/on of the hadronic contributions will rather be taken care
of by the relative strength of the temperature dependent hadronic masses
to the quark masses.

The temperature dependent mesonic masses are obtained from the pole
condition
\begin{equation} 
1-2G_M\Pi_M(\omega=m_M,\vec{k}=0)=0.
\label{eq.Pole} 
\end{equation}
Here $G_M$ is the effective vertex factor for the given flavor
combination and $\Pi_M(k^2)$ is the one-loop polarization function for
corresponding mesonic channel given by the Random Phase
Approximation~\cite{Fetter} as,
\begin{equation}
 \Pi_M(k^2)=\int \frac{d^4p}{(2\pi)^4}Tr[\Gamma_M S(p+\frac{k}{2})
 \Gamma_M S(p-\frac{k}{2})],
\label{eq.pimk}
\end{equation}
where $S(p)$ is the quark propagator. In this work we shall only
consider the lowest lying nonet mesons. The details of the calculations
may be found in our earlier work~\cite{Datta,Fin}. The final
computations however consider the reparametrized PNJL model as discussed
in the previous section. The mesonic contribution to the thermodynamic
potential is given as~\cite{Robner},

\begin{equation}
 \delta\Omega_M=-\nu_M T\int \frac{d^3p}{(2\pi)^3}
 {\rm ln}(1-e^{-\frac{E_p}{T}})
\label{eq.mesmas}
\end{equation}
where $\nu_M$ is statistical weight factor of corresponding mesonic
species and $E_p=\sqrt{\vec{p}^2+m_{pole}^2(T)}$, where $m_{pole}$ is
the mesonic mass obtained by solving Eq.(\ref{eq.Pole}). 

In the baryonic sector, the lower lying window is occupied by the
nucleons, protons and neutrons. They having a bare mass $\sim$ 940 MeV
contribute insignificantly to the thermodynamics. Also chiral
perturbation theory results like in \cite{Leutwyler} indicate that the
nucleon masses increase with temperature apart from a very small decrease
in the intermediate regimes of temperature. In this study we therefore
consider only the constant mass for nucleons. Role of other baryon
species are left out in this excercise. The baryonic contribution to
the thermodynamic potential is given by~\cite{Sasaki},

\begin{equation}
 \delta\Omega_B= \nu_B T\int\frac{d^3p}{(2\pi)^3}
 {\rm ln}(1+e^{-\frac{E_p}{T}})
\label{eq.barmas}
\end{equation}
where $\nu_B$ is statistical weight factor of corresponding baryonic
species. The final thermodynamic potential is the sum of the parts
obtained from Eq.(\ref{eq.Potential}), Eq.(\ref{eq.mesmas}) and
Eq.(\ref{eq.barmas}). We shall refer to this as the Modified Polyakov
loop enhanced Nambu$-$Jona-Lasinio (MPNJL) model.

\section{Results}

\begin{figure}[!htb]
{\includegraphics[height=6cm,width=8.0cm,angle=360]
{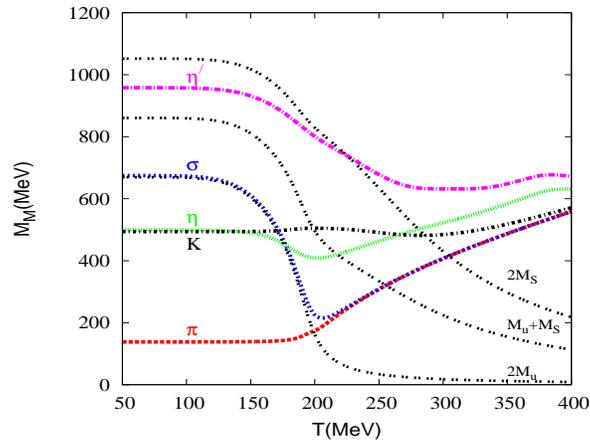}}
\caption{(color online)Pseudoscalar and scalar meson masses plotted as a
function of temperature.} 
\label{fg.mass}
\end{figure}

The thermodynamic potential \ref{eq.Potential} is minimized with respect
to the $\sigma$, $\Phi$ and $\bar{\Phi}$ to obtain the mean fields.
These are then inserted into the equations \ref{eq.Pole} and
\ref{eq.pimk} to obtain the meson masses as function of temperature.  In
Fig.~\ref{fg.mass} we have plotted the variations of meson masses for
$\pi$, $\sigma$, $K$, $\eta$ and $\eta^{\prime}$ with temperature.
The results are similar to some of the earlier works \cite{Datta,
Costa}.  The most significant change is the mass of the $\pi$ which
rises from about 140 MeV near the crossover temperature to 550 MeV as
the temperature nears 400 MeV. The $\sigma$ mass has the expected
behavior of first a strong decrease to reach the $\pi$ mass and then
increase along with the $\pi$ mass. In this temperature range the
$\eta^{\prime}$ mass decreases by almost 300 MeV. The masses of the $K$
and $\eta$ vary by a relatively small amount. The various combinations
of constituent quark masses are also plotted for comparison. Obviously
the signature of the chiral symmetry restoration in the meson sector at
high temperatures is evident as the constituent masses go down.

\begin{figure}[!htb]
\subfigure[]{\includegraphics[height=6cm,width=8.0cm,angle=360]
{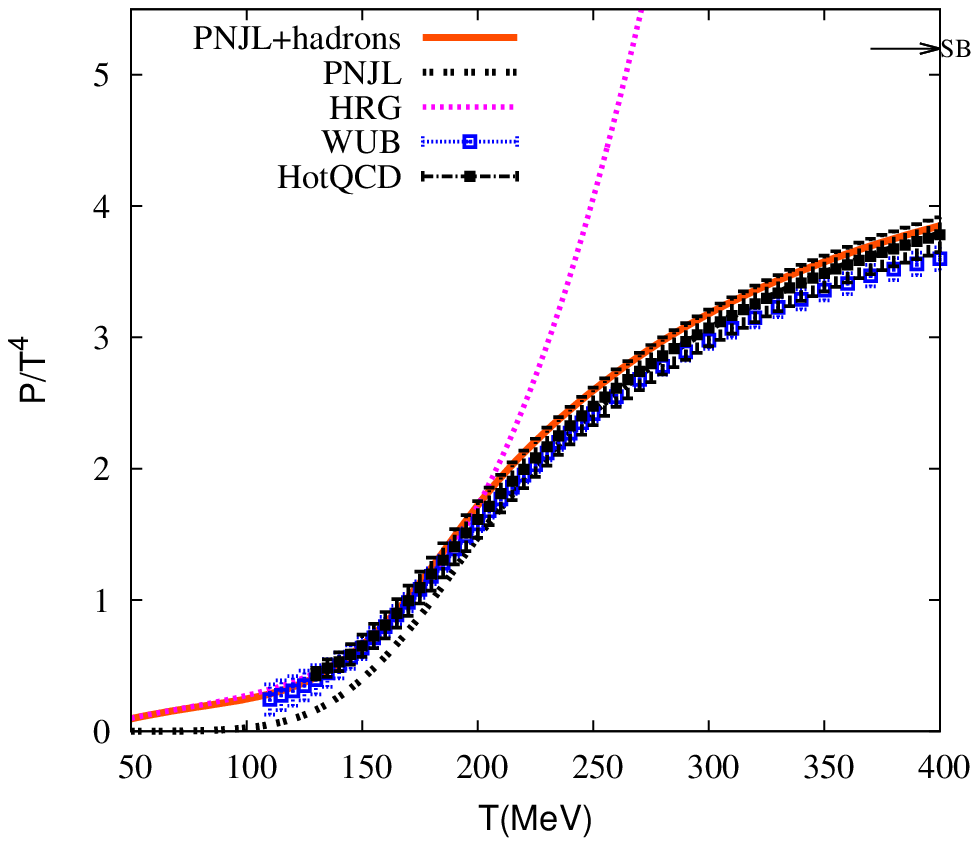}
\label{fg.pressure}}
\subfigure[]{\includegraphics[height=6cm,width=8.0cm]{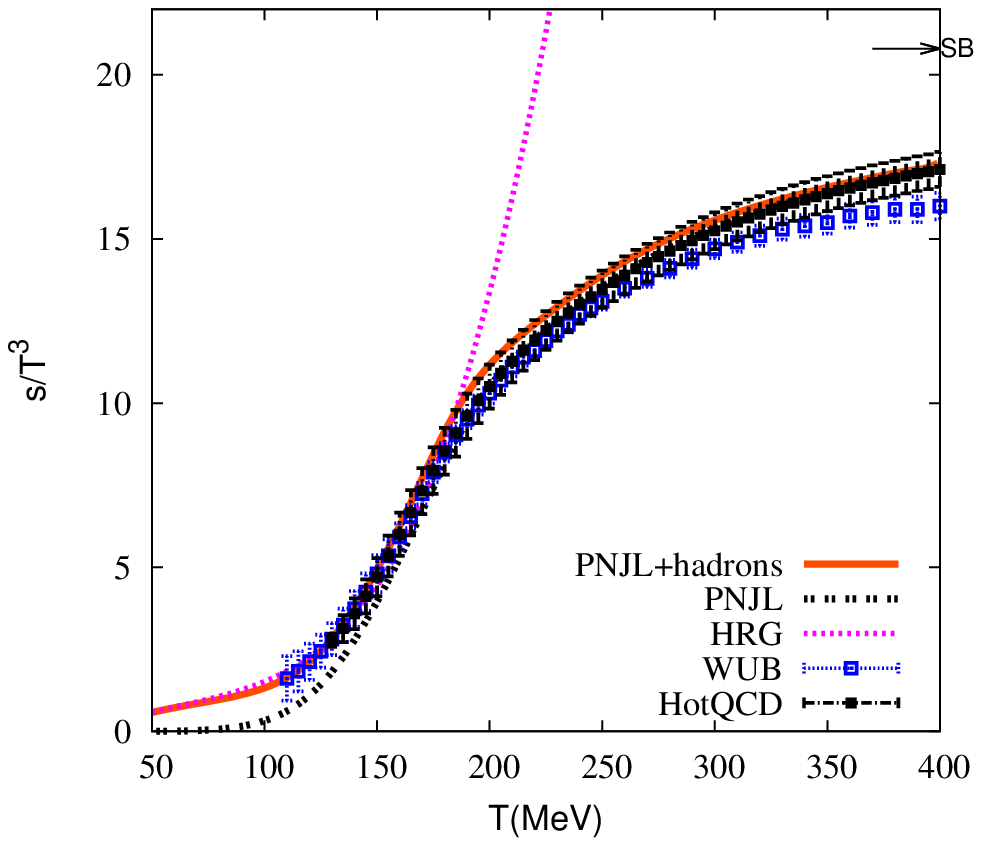}
\label{fg.entropy}}
\caption{(color online) Scaled pressure and entropy plotted as
functions of temperature.} 
\end{figure}

The mean fields are then put back in Eq.(\ref{eq.Potential}), and the
pole masses in the Eq.(\ref{eq.mesmas}) and Eq.(\ref{eq.barmas}) to
obtain the pressure. The scaled pressure and scaled entropy density are
shown in Fig.~\ref{fg.pressure} and Fig.~\ref{fg.entropy} respectively.
These are compared to the continuum extrapolated lattice QCD
(HotQCD~\cite{Bazavov14}, and Wuppertal-Budapest~\cite{Borsanyi14})
data. Both the quantities in the MPNJL model agree with the usual PNJL
model and lattice QCD data for the higher temperatures. At the lower
temperature the MPNJL model remarkably reproduces the lattice QCD data,
where the PNJL model fails. Obviously a similar result was obtained with
the hybrid PNJL model~\cite{hybrid}.  But unlike the hybrid model, where
the switching function had to be tuned, here we have no extra
parameters, apart from those already present in the PNJL model.

Given that we have considered only a few mesons corresponding the flavor
SU(3) octet and the lowest lying nucleons, the agreement of the bulk
thermodynamics in the MPNJL model and lattice QCD data is surprising.
However as shown in the figures, the scaled pressure and scaled entropy
density obtained in the ideal Hadron Resonance Gas model have an
excellent overlap with the MPNJL and lattice QCD data in the low
temperature region. It therefore seems sufficient to include the limited
number of hadrons for the present study.

\begin{figure}[!htb]
\subfigure[]{\includegraphics[height=6cm,width=8.0cm]
{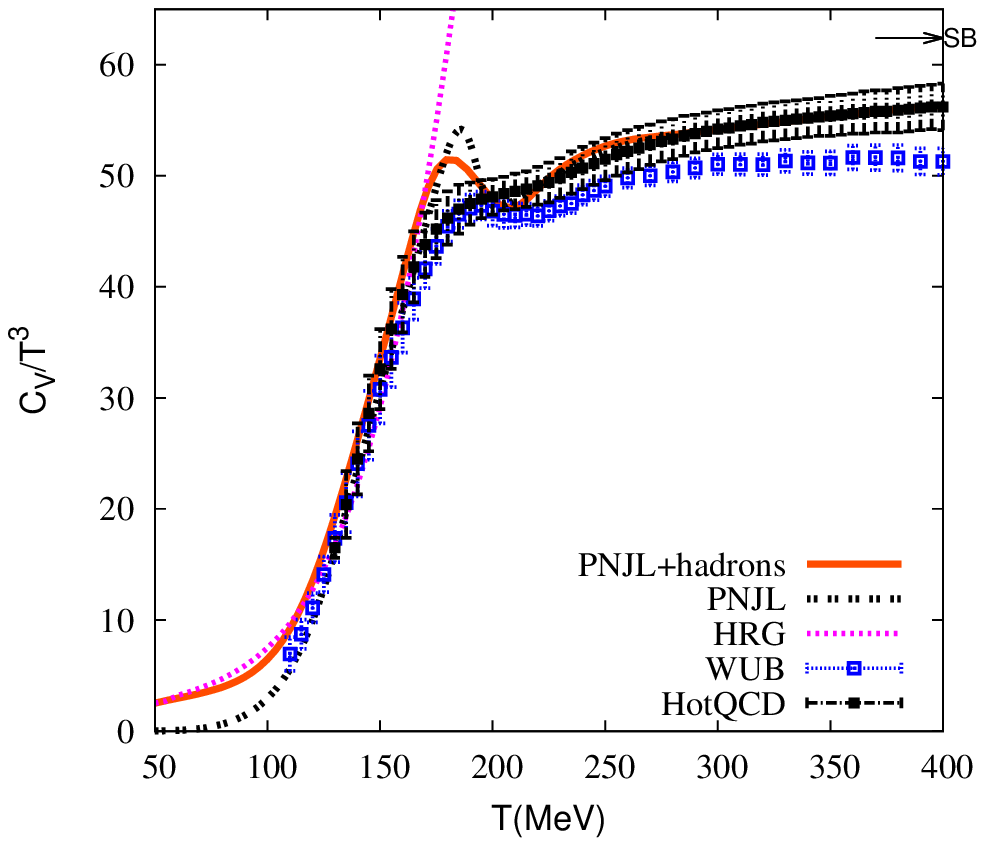}
\label{fg.cv}}
\subfigure[]{\includegraphics[height=6cm,width=8.0cm]
{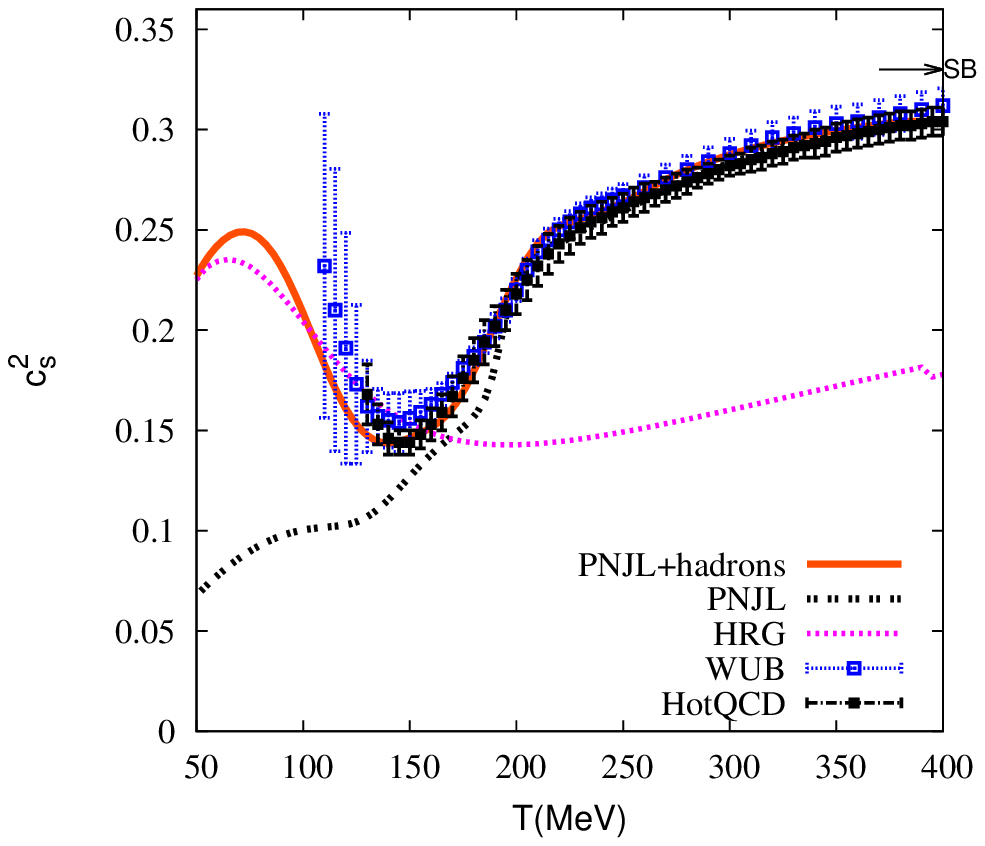}
\label{fg.cs2}}
 \caption{(color online) Specific heat and speed of sound as 
functions of temperature} 
\label{cvcs2}
\end{figure}

With the MPNJL model we now obtain the specific heat at fixed volume,
which includes the second derivative of the thermodynamic potential with
respect to temperature. The variation of the scaled specific heat with
temperature is shown in Fig.~\ref{fg.cv}. For higher temperatures the
lattice results are distinctly different. In fact the difference can be
seen to be gradually increasing as we move from the scaled pressure to
scaled entropy density to finally the scaled specific heat. As already
mentioned in \cite{repara}, we chose the parameters in the Polyakov loop
potential to agree with the HotQCD data. In the lower temperature ranges
it is difficult to conclude if the results of the MPNJL model may be
preferred over the PNJL model results when compared to the lattice QCD
data.

To bring out the difference between the two models we therefore consider
the squared speed of sound, which turns out to be the ratio of the
entropy to the specific heat at fixed volume. This is shown in
Fig.~\ref{fg.cs2}. Here we see a wide difference between the PNJL and
MPNJL model results for the lower temperatures. The MPNJL model results
indeed agrees well with both the Hadron Resonance Gas model, as well as
the lattice QCD data.

With these results we demonstrated the necessity and utility of
introducing the beyond mean field contributions to the PNJL model,
though with quite a few assumptions. As we see that here we did not need
any extra switch between the hadron and PNJL contributions. The switch
between the degrees of freedom are affected by the varying masses with
temperature. For lower temperatures the constituent quark masses are
quite high and the meson masses comparatively low, giving rise to meson
domination. This is in addition to the suppression of quark excitation
by the Polyakov loop. The condition is reversed as one approaches higher
temperatures, and the system becomes quark dominated.

\begin{figure}[!htb]
 {\includegraphics[height=6cm,width=8.0cm]
{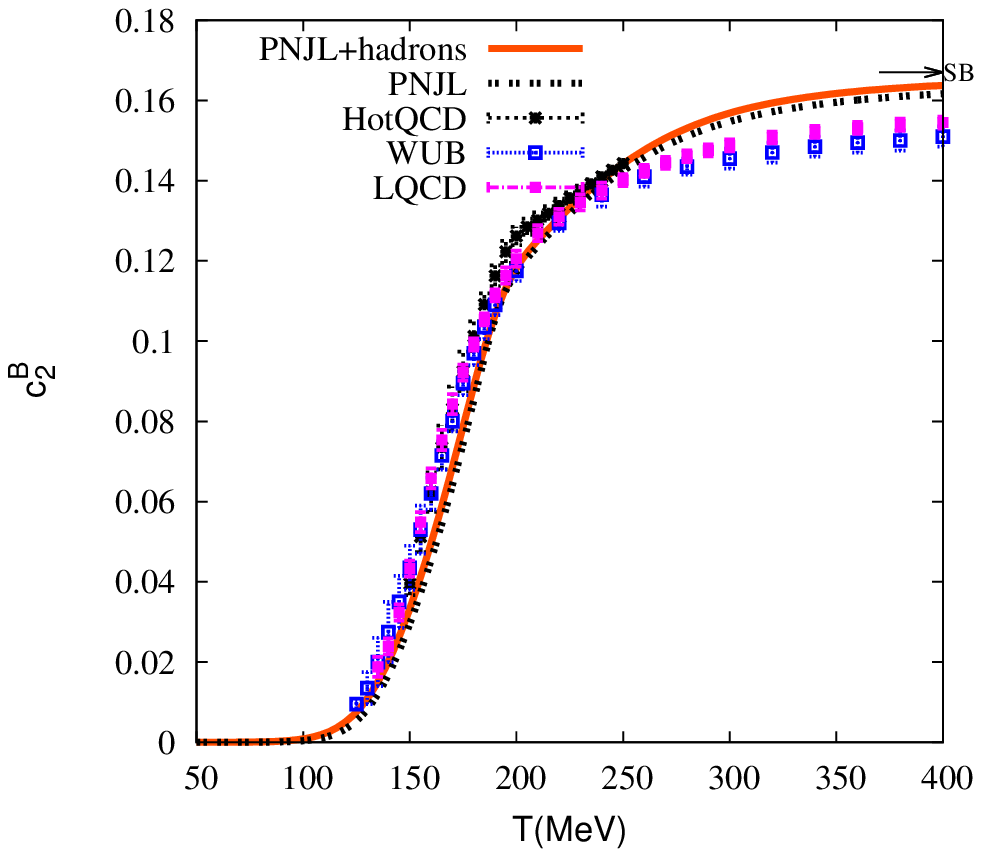}}
{\includegraphics[height=6cm,width=8.0cm]
{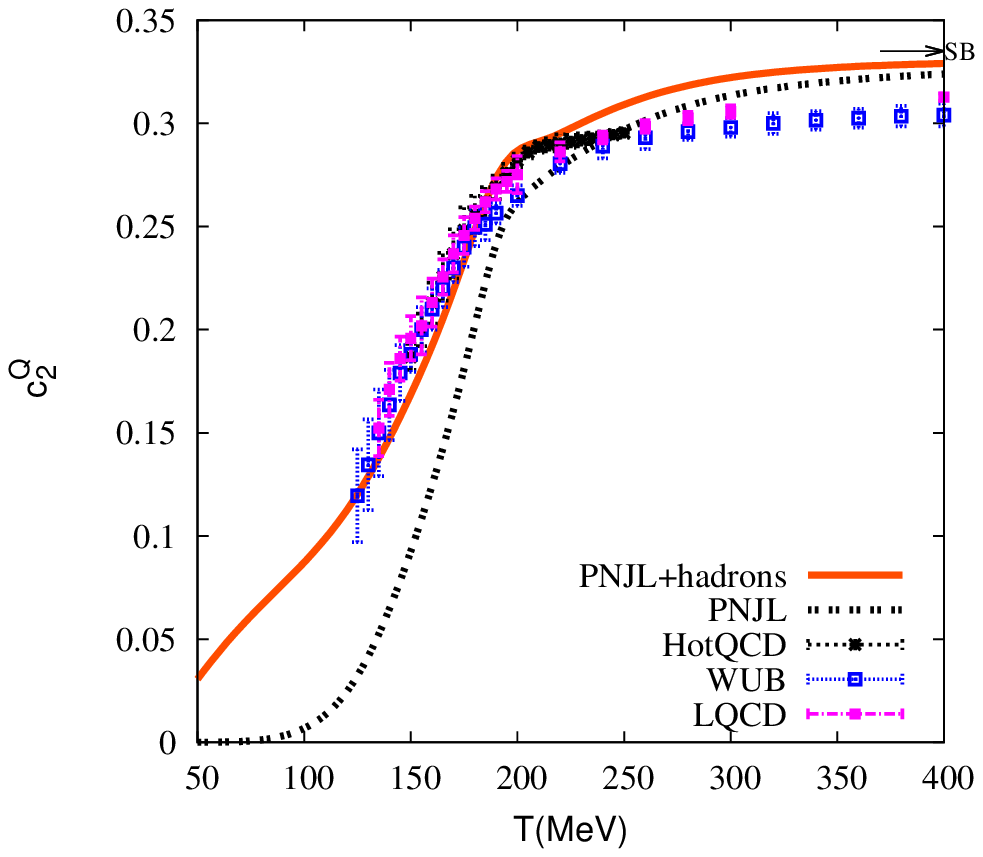}}
{\includegraphics[height=6cm,width=8.0cm]
{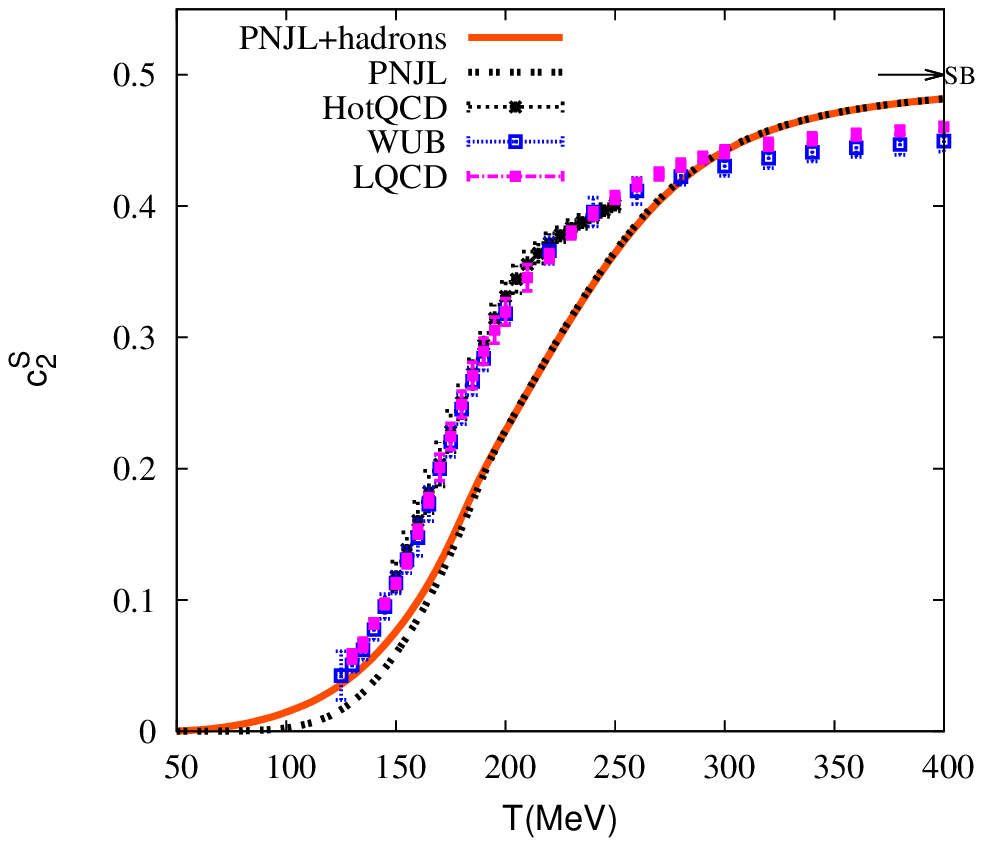}}
 \caption{(color online) Conserved charge fluctuations as 
functions of temperature} 
\label{fg.fluct}
\end{figure}

A proper determination of the state of strongly interacting matter at
finite temperatures and chemical potentials requires knowledge of the
fluctuations of conserved charges~\cite{EjiriKarsch, Deb, Lahiri,
BAbelev}. They also act as indicators of phase transition or crossover
through which the system passes \cite{HattaStephanov, Saha, Jeon, Koch,
Asakawa, BhatDas, Fin, BhatRay, BhatSamanta, SahaUpadhaya,
StephanovRajagopal}.  At a given temperature and arbitrary chemical
potentials, the pressure of the system may be expanded as a Taylor
series around zero chemical potentials, where the coefficients are
directly related by the fluctuation-dissipation theorem
\cite{GhoshLahiri} to the fluctuations at various orders. The n-th order
Taylor expansion coefficient $c_n^X(T)$ of scaled pressure can be
written in terms of fluctuations $\chi_n^X(T)$ of conserved charges
(baryon number $B$, electric charge $Q$ and strangeness $S$) as,

\begin{equation}
c_n^X(T)=\frac{1}{n!}\frac{\partial^n(P/T^4)}{\partial(\mu_X/T)^n}=
\frac{T^{n-4}}{n!}\chi_n^X(T)
\end{equation}
where the expansion is carried out around $\mu_B=\mu_Q=\mu_S=0$. In
Fig.~\ref{fg.fluct} we present our results for the second order
fluctuations of the conserved charges along with a comparison of the
continuum data from lattice QCD \cite{Bazavovfluct, BorsanyiFodor,
BellwiedBorsanyi}.  In the model, these fluctuations are obtained by a
suitable Taylor series fitting as discussed in detail in \cite{Ray}.

The baryon number fluctuation $c^B_2$ obtained in the PNJL and the MPNJL
model are very close to each other in the whole range of temperatures
studied.  The only hadrons that contribute additionally in the MPNJL
model are the nucleons with a mass $\sim$ 1 GeV, which is much heavier
than the corresponding constituent mass of the quarks. So the difference
between the two model results is insignificant. There is a possible
concern for overcounting the baryons in the MPNJL model $-$ as the
constituent quarks and as the nucleons. Obviously this would be of concern
as more and more baryons are included. But in the present case the
constituent quarks overwhelm the system due to both their lower masses
as well as larger degrees of freedom. In view of these discussions, it
is surprising to find a reasonable agreement of the results from the
PNJL model with the lattice QCD data even for temperatures below 150
MeV. It seems that the partonic fluctuations manifest themselves
strongly in the baryon susceptibilities.

This is not the case for the electric charge fluctuations $c^Q_2$. The
PNJL and MPNJL model results differ significantly for temperatures close
to 200 MeV. The MPNJL model has a very good agreement with the lattice
QCD results. The dominant hadrons in this sector are the pions. For very
low temperatures the pion mass is almost half the mass of the
constituent quarks. With increase in temperature the pion mass increases
and the quark mass decreases such that the combination nicely reproduces
the lattice QCD data.

The strangeness fluctuation $c_2^S$ in the PNJL and MPNJL models differ
for low temperatures by a smaller amount when compared to the charge
fluctuations. For the lowest temperatures in the lattice QCD data, the
MPNJL model seems to agree. But thereafter the two models merge and they
deviate from the lattice QCD data. In the hadronic sector the dominant
contributors are the $K$ and $\eta$. Their masses, though almost half of
the constituent masses $M_u+M_s$, are still quite large. Moreover the
$K$ mass is almost constant over the whole temperature range. On the
other hand the decrease in the constituent mass of the strange quark
with temperature is not fast enough. So their contributions to
strangeness fluctuations above temperatures of 150 MeV is not enough to
agree with the lattice QCD data.

\section{Conclusion}

Numerous attempts are being made to predict the correct EoS for the
strongly interacting system. Lattice QCD is the most robust abinitio
technique. However effective models that are much easier to handle and
suitable enough to extract interesting physical outcomes are regularly
employed. The reliability of such models in quantitative estimates have
often come under review. In this regard we are investigating the
various possible improvements for the PNJL model so that it can serve as
an effective tool for quantitative analysis of strong interactions in
chemical equilibrium.

In an earlier work~\cite{repara}, the Polyakov loop potential was
reparametrized to bring various thermodynamic quantities in reasonable
agreement with the lattice QCD data. Among the issues pointed out in
that work was the insufficiency of the PNJL model to reproduce the
correct results for temperatures close to and below the crossover
transition. The relevance of the hadronic degrees of freedom was
realized and a hybrid model was built~\cite{hybrid} with the HRG model
and PNJL model coupled via a switching function. The method worked well,
but a more natural framework was sought. The existing literature on the
beyond-mean field calculations in the NJL and PNJL models led us to
propose the present modified PNJL model, where the hadronic
contributions are additively included. There is no switching function,
but the hadrons are given medium modified masses. The relative variation
of the hadron and constituent quark masses with temperature effectively
selects the dominant degrees of freedom. The best utility of this scheme
over the scheme using switching function is for finite chemical
potentials. The parameters in the switching function being additional
parameters had to be fixed at various temperatures and chemical potentials.
Here on the other hand there are no extra parameters and the effect of
temperature and chemical potentials are taken care of through the
respective hadronic distribution functions. 

The scheme is found to satisfactorily reproduce the lattice QCD results
for a range of observables including the pressure, entropy, specific
heat, speed of sound and the baryon number and electric charge
fluctuations. The results from the model however deviated significantly
from the lattice data for strangeness fluctuations. To address this
shortfall it is necessary to revisit the strangeness sector of the PNJL
model, which we hope to address in future.

\section{Acknowledgement}
The authors would like to thank DST, DAE, Board of Research in 
Nuclear Sciences (BRNS), University Grants Commission (UGC). 
AB thanks Alexander von Humboldt (AvH) foundation and Federal 
Ministry of Education and Research (Gernamy) for support 
through Research Group Linkage programme. K.S. acknowledges DST-SERB for
financial support under NPDF file no. PDF/2017/002399. S.M. thanks CSIR for 
financial support. K.S. and S.U. would like to thank Souvik Priyam Adhya 
for useful discussions.


\end{document}